\documentstyle[preprint,12pt,aps,pra]{revtex}
\input epsf
\tightenlines

\title{Calculation of the energy spectrum of a two-electron spherical 
quantum dot}

\author{Ramiro Pino  \footnote{E-mail: rpino@win.tue.nl}} 
\address{Department of Mathematics and Computer Science, Technische 
Universiteit Eindhoven,\\  P.O. Box 513, Eindhoven 5600 MB, The Netherlands.}
\author{V\'{\i}ctor M. Villalba \footnote{E-mail: villalba@pion.ivic.ve}}
\address{Centro de F\'{\i}sica,
Instituto Venezolano de Investigaciones Cient\'{\i}ficas,\\
Apartado 21827, Caracas 1020-A, Venezuela.} 

\begin{document}

\maketitle

\begin{abstract}
We study the energy spectrum of the two-electron spherical parabolic
quantum dot using the exact Schr\"odinger, the Hartree-Fock, and the
Kohn-Sham equations. The results obtained by applying the shifted-1/N 
method are compared with those obtained by using an accurate numerical 
technique, showing that the relative error is reasonably small, although 
the first method consistently underestimates the correct 
values. The approximate ground-state Hartree-Fock and local-density 
Kohn-Sham energies, estimated using the shifted-1/N method, 
are compared with accurate numerical self-consistent solutions.
We make some perturbative analyses of the exact energy in terms of the
confinement strength, and we propose some interpolation formulae. 
Similar analysis is made for both mean-field approximations and 
interpolation formulae are also proposed for these exchange-only 
ground-state cases.
\end{abstract}


\section{Introduction}

Progress in nanotechnology has allowed the development of small devices 
like quantum dots. The confinement potential can be safely approximated 
by a harmonic one \cite{sime,peet}, which has boosted the study of quantum 
dots with parabolic confinement during the last years (see e.g. Refs. 3 
and 4, and references therein). The presence of many interacting electrons 
render the computation of the electronic states and properties a very 
complicated many-body problem. The first non-trivial exactly solvable 
problem of many-electrons is the one of two electrons confined in a 
parabolic potential, which made it a very attractive workbench for testing 
all kind of approximations (see e.g. \cite{KHL,LF,DN,ZLYOK,ES1,MiBe}). 

The Hamiltonian describing the pair of interacting electrons in a parabolic 
quantum dot in the effective masss approximation can be written as
\begin{equation}
\hat{H}=-\frac{\hbar^2}{2m^*} ( \nabla_1 ^2 + \nabla_2 ^2 ) +
\frac 1 2 m^* \omega^2 ( r_1 ^2 + r_2 ^2)  +
\frac{e^2}{\epsilon |\vec r_1 - \vec r_2|}
\label{eq1}
\end{equation}
where $ m^{*}$ is the effective mass, $\omega$ the confinement strength, 
and $\epsilon $ the dielectric constant of the
host material and $\nabla ^2$ is the Laplacian operator.

Equation (\ref{eq1}) can be separated into a centre-of-mass and a relative
motion component. Furthermore, due to the radial symmetry of the components,
only those parts of the corresponding Schr\"odinger equations have to be
solved. The centre-of-mass part will give the well-known harmonic 
oscillator problem. In three dimensions, the relative motion part may
admit exact solutions for special choices of the parameters (see e.g.
\cite{KHL}).
For the two dimensional case, similar separation and substitution can be 
made, and again there are no general solutions expressible in terms of 
special functions. Nevertheless, in Ref. \cite{Lozanskii} and more 
recently in Ref. \cite{Taut2} it was shown that there exist analytic 
solutions for special choices of the confinement constant. 

Many-body effects are usually divided into exchange and correlation 
components \cite{PY}. Exchange-only effects are considered in Hartree-Fock
(HF) and differently, in Kohn-Sham (KS) approaches (although in KS, the
correlation effects can be included), which typically amounts for around
$10\%$ of the total energy. Correlation is about one order of magnitude
smaller. Nevertheless, it has been shown that in two dimensions for two 
electrons
in a harmonic field, correlation may play a bigger role specially for
singlet states (see e.g. \cite{PGM}).

The shifted-1/N method \cite{IPS,IS} has been applied to various 
condensed matter problems. Also, the two-dimensional relative motion
Schr\"odinger problem have been solved using this technique
in Refs. \cite{ES1,ES2}.

The article has been structured as follows: in Sect II we describe the
Schr\"odinger, Hartree-Fock, and Kohn-Sham approaches we use. In Sect. III
we solve the exact, HF, and KS-LDA equations using an accurate numerical 
technique and the shifted-1/N method. We also apply perturbation theory
up to first order in both limits of confinement for the exact and
mean field cases, and we propose some interpolation formulae for 
the energy. We discuss the accuracy of the mean field approaches, and
of the shifted-1/N method for the present case.

\section{Method}

Throughout the paper the units of energy will be given in terms of the 
effective Rydberg constant ${\cal R}^{*}=\hbar ^2/(2 m^* a^{*2})$ and the 
effective Bohr radius $a^{*}=\hbar ^2\epsilon /m^{*}e^2,$ respectively.

In centre-of-mass and relative coordinates and measuring in reduced units
the Hamiltonian reads 
\begin{equation}
\hat{H}=-(\frac 1 4 \nabla_{\vec R} ^2 + \nabla_{\vec r} ^2 ) +
\gamma ^2 R^2 + \frac{\gamma ^2} 4 r^2 + \frac 1 r
\label{eq2}
\end{equation}
where we have chosen the centre-of-mass $\vec R=(\vec r_1 + \vec r_2)/2$ 
and the relative coordinate $\vec r=(\vec r_1 - \vec r_2)$.

The separation leads to the harmonic oscillator problem for the 
centre-of-mass coordinate, with energy
\begin{equation}
E_{NL} = \gamma \; ( 2N + L + 3/2 ) \; .
\label{eq3}
\end{equation}
and eigenfunctions
\begin{equation}
\Psi_{NLM}(\vec R)= { \cal N}_{NL} \exp( - \gamma R^2 ) (2 \gamma)^{L/2} 
R^L L_N^{L+1/2}( 2 \gamma R^2 ) Y_{LM}(\theta_R,\phi_R) \; .
\label{eq4}
\end{equation}

For the relative coordinate equation the wavefunction can be separated into
radial and angular components
\begin{equation}
\Psi(\vec r) =\frac{u(r)}{r} Y_{lm}(\theta, \phi) \; ,
\label{eq5}
\end{equation}
where $Y_{lm}(\theta, \phi)$ are the spherical harmonics which are 
eigenfunctions of the angular momentum operator $L_z$, and $L^2$ 
with eigenvalues $m$ and $l$. This separation makes the 
corresponding radial Schr\"odinger equation the following second order 
ordinary differential equation
\begin{equation}
\left[ -\frac{d^2}{dr^2}+l(l+1)\frac 1{r^2} + \frac 14 \gamma ^2 r^2 +
\frac 1r - E \right] u(r) = 0 \; .
\label{eq6}
\end{equation}
It is well known that exact solutions of equation (\ref{eq6}) cannot be
expressed in a closed form in terms of special functions. There are
analytic expressions for the energy for particular values of $\gamma$ and
$l$ as it was pointed out in Refs. \cite{KHL,KeSi}, among others.

The electrons should satisfy the Fermi-Dirac statistics which means in this
case that for singlet states ($s=0$) the spatial part of the wavefunction 
should be antisymmetric and for triplet states ($s=1$) symmetric. As the
centre-of-mass coordinate remains the same after exchanging to electrons,
the antisymmetry requirement will be in the relative part. Because
of the separation in radial and angular components of the 
relative-coordinate wavefunction it will mean that singlet states are 
associated with odd $l$ and triplet states with $l$ even, respectively.

It is interesting to compare the results of exact calculations with 
independent-electron models like Hartree-Fock (HF), and Kohn-Sham (KS) 
\cite{PY} in order to assess the relative importance of many-body 
effects like exchange and correlation, and also to evaluate the performance
of the local-density approximation. For two paired electrons the
electronic density is $\rho_{HF} = 2 |\phi_{HF}|^2$, where $\phi_{HF}$
is the orbital, and the exchange potential is equal to half of the Hartree 
one with opposite sign. The HF equation can be written as
\begin{equation}
 \left[ - \frac 12 \nabla^2 + v(r) + \frac 12 v_H[\rho_{HF}] \right] 
\phi_{HF} = \varepsilon_{HF} \: \phi_{HF} \; ,
\label{hf1}
\end{equation}
$v(r)=\frac 12 \gamma^2 r^2$ and
$\varepsilon_{HF}$ is the HF orbital energy. The total HF energy is 
written as
\begin{equation}
E_{HF} = 2 \varepsilon_{HF} - \frac 12 \int d \vec r \: \rho_{HF} \: 
v_H [\rho_{HF}] \; ,
\label{hf2}
\end{equation}
where $v_H$ is the Hartree potential given by
\begin{equation}
v_H [\rho] = \int d \vec r^{\; \prime}
\frac{\rho(\vec r^{\; \prime})}{|\vec r - \vec r^{\; \prime}| } \; .
\label{hartree}
\end{equation}
The KS equation can be written as
\begin{equation}
\left[ -\frac 12 \nabla^2 + v(r) + v_H[\rho_{KS}] + v_x(\rho_{KS}) \right] 
\phi_{KS} = \varepsilon_{KS} \: \phi_{KS}
\label{ks1}
\end{equation}
$\varepsilon_{KS}$ and $\phi_{KS}$ are the KS orbital energy and 
eigenfunction, respectively, and again $\rho_{KS} = 2 |\phi_{KS}|^2$.
We take here the local-density
approximation for which $v_x(\rho)= \frac 43 c_x \rho^{1/3}$, with
$c_x =-\frac 34 \left( \frac{3}{\pi} \right)^{1/3}$ (see e.g. \cite{PY}).
The total KS energy is given by
\begin{equation}
E_{KS} = 2 \varepsilon_{KS} - \frac 12 \int d \vec r \: \rho_{KS} \: 
v_H [\rho_{KS}] - \frac 13 \int d \vec r \: \rho_{KS} \: v_x (\rho_{KS}) \; .
\label{ks2}
\end{equation}
Eqs. (\ref{hf1}) and (\ref{ks1}) have asymptotae controlled by
the harmonic potential, so the asymptotic density looks like
\begin{equation}
\rho_a(\vec r) \propto \exp( -\gamma r^2 )  \; .
\label{ks3}
\end{equation}

\section{Results and Discussion}

First we make an analysis of the solution of the exact case using
the 1/N approximation.  For details in the derivation of the formulae 
related to the shifted-1/N method, we 
refer to the literature (see e.g. \cite{ES1,IPS,IS,ES2}).  

In order to establish the accuracy of the results obtained
by the application of the shifted $1/N$ method we compare them with 
those obtained by using the Schwartz's numeric method \cite{Schwartz}. 
The method is based in a numerical approximation of functions on a mesh
and gives very accurate results \cite{vp,pv}.  
There are only empirical estimates of the error \cite{Schwartz} which
turns out to be exponentially decaying with the number of points given 
the mesh step.  
The interpolation function is chosen as
\begin{equation}
f(r)=\sum_{m}f_{m}\frac{u(r)}{(r-r_{m})a_{m}},
\end{equation}
where 
\begin{equation}
u(r)=\sin [\pi (r /h)^{1/2}].
\end{equation}
Here $r_m$ is a zero of $u(r)$, $a_m$ is a zero of its derivative, and 
$h$ is the step of the mesh which turns out to be quadratically spaced.
The choice of the step $h$ was made after estimating the characteristic
length of the effective potential, and then multiplying the obtained
estimate by five and dividing it by the square of the number of points in 
the mesh, usually around 300. This guarantees that $h$ is minimal for a 
given $\omega$, and also that the function value at the last mesh point 
is practically zero.  

In figure 1 we show the behavior of the error of the energy of few lowest 
eigenvalues calculated using the shifted-1/N method, compared with the 
accurate results obtained using the Schwartz's method. We plotted the 
relative error defined as $\delta E = E_{approx}/E_{exact}-1$, as function
of the reduced variable $\gamma^{\; \prime}=\gamma /(\gamma +1)$.  
The $0s, 0p$, and $0d$ states are the three lowest energy states of the
relative motion with angular momentum $l=0, 1$, and $2$, respectively. 
It can be seen that the error remains bounded in $0.5 \%$ for the first
eigenvalue, in $0.075\%$ for the second one and $0.022\%$ for the third
one. It is noticeable that the method always underestimates the correct
values of the energy, and that the error decreases with the increase of 
the angular
quantum number $l$, as it should be expected. Furthermore, the error has
some maximum between the two limiting cases, after which it decreases as
expected, since the shifted-1/N method reproduces exactly the oscillator
case.

Figure 2 shows the behavior of the total energy (the sum of the 
centre-of-mass
and relative motion energies) as a function of $\gamma ^{\; \prime}$, as
calculated using the shifted-1/N method. The first two symbols are the
indices of the centre-of-mass component of the energy, and the last two 
correspond to the relative motion. The lowest six states are depicted.
The inset is a magnification  for $\gamma^{\; \prime}$ from
$0.01$ to $0.15$. Here, an apparent linear behavior can be observed, but 
a more careful analysis indicates that, for instance, for the ground-state
first six-eight points, the
effective power in terms of $\gamma ^{\; \prime}$ is about $0.8$. In the
last few points the curve can be fitted well with a function of the
type $\gamma^{\; \prime}/(1-\gamma^{\; \prime})$ which is the expected
behavior for electrons in a harmonic field.
The relative error is not shown, but it is estimated in roughly one half 
of the one shown in figure 1, since the centre-of-mass energy can be 
calculated exactly and it is typically of the size of the relative
motion energy. 

Although we are able to solve the Schr\"odinger equation very accurately
for this system, perturbation analysis can give some more insight of
the behavior of the electrons under weak and strong confinement.
For weak confinement ($\gamma \rightarrow 0$) we have that the kinetic
energy term can be neglected (see e.g. \cite{Taut2,Gonz}), which corresponds 
to the strong interaction limit (Wigner crystallization). Then, the energy
is approximately taken as the minimum of the effective potential (this is 
the zeroth order approximation to the energy), 
\begin{equation}
V_l(r) = l(l+1)\frac 1{r^2} + \frac 14 \gamma ^2 r^2 + \frac 1r  \; ,
\label{eq13}
\end{equation}
the minimum is reached for $r_0^l$ that satisfies the equation 
\begin{equation}
\gamma^2 r_0^{l \; 4} - 2 r_0^l - 4 l (l+1) = 0 \; .
\label{eq13a}
\end{equation}
For $l=0$ $r_0^0 = (2/\gamma^2)^{1/3}$ and for $l \gg \gamma/2$
$r_0^l \approx (2 l/\gamma)^{1/2}$. The minimum of the potential is then
\begin{equation}
U_0^l = l(l+1)\frac{1}{r_0^{l \; 2}} + \frac 14 \gamma ^2 r_0^{l\; 2} + 
\frac{1}{r_0^l}  \;  ,
\label{eq13b}
\end{equation} 
which is $2^{-1/3} 3/2 \gamma^{2/3}$ for $l=0$ and $\gamma (2 l+1)/2 + 
\gamma^{1/2}/(2l)^{1/2}$ for large enough values of $l$.

The next order in the approximation is to get the effecti\-ve frequen\-cy 
$\gamma_l^2 = \frac 12 d^2 V_l(r)/dr^2 |_{r=r_0^l}$,
\begin{equation}
\gamma_l^2 = \frac 14 \gamma^2 + \frac{1}{r_0^{l \; 3}} +
\frac{3 l (l+1)}{r_0^{l \; 4}}  \; ,
\label{eq13c}
\end{equation} 
which is $\gamma_0^2 = 3/4 \gamma^2$  and $\gamma_l^2 =
\gamma^2 + 3/4 \gamma^2/l +2^{1/2}/4 (\gamma/l)^{3/2}$ for large $l$.
Now we can estimate the energy levels of this effective harmonic field. So 
the total energy in the weak confinement limit is
\begin{equation}
E_{nl} \approx l(l+1)\frac{1}{r_0^{l \; 2}} + \frac 14 \gamma^2 r_0^{l\; 2}
 + \frac{1}{r_0^l} + 2 \gamma_l ( n + 1/2 ) \; ,
\label{eq14}
\end{equation}
where $n=0,1,..$. For $l=0$
\begin{equation}
E_{n0} \approx  \frac 32 2^{-1/3} \gamma^{2/3}
+  \frac{3^{1/2}}{2} \gamma ( n + 1/2 ) \; .
\label{eq14b}  
\end{equation}
For $l$ large enough 
\begin{equation}
E_{nl} \approx \frac 12 \gamma (2 l+1) + \left( \frac{\gamma}{2l} 
\right)^{1/2} + 2 \left[ \gamma^2 + \frac 34 \frac{\gamma^2}{l} + 
\frac{2^{1/2}}{4} \left( \frac{\gamma}{l} \right)^{3/2} \right]^{1/2} 
( n + 1/2 ) \; .
\label{eq14c}
\end{equation}
Equation (\ref{eq14b}) gives the explanation why the effective
power of the ground-state energy for small $\gamma^{\; \prime}$ is
approximately $0.8$: it is between $2/3$ and $1$, the effective
powers for the weak field limit for the relative motion and 
the centre-of-mass energies in terms of $\gamma^{\; \prime} \approx
\gamma$.

In the strong confinement regime, 
the zeroth order approximation amounts to neglect the electron-electron
interaction, so it corresponds to the oscillator's energy.
Application of the first order of the perturbation theory \cite{Dav} for 
strong confinement ($\gamma \rightarrow \infty$) for the relative coordinate 
equation leads to 
\begin{equation}
E_{nl} \approx E_{nl}^{osc} + \langle \phi^{osc}_{nl} | r^{-1} | 
\phi^{osc}_{nl} \rangle \; ,
\label{eq15}
\end{equation}
where $ E_{nl}^{osc} = 2 \gamma ( 2 n + l + 3/2) $ and substituting 
$\phi^{osc}_{nl}$ (which is similar to equation (\ref{eq4}) ) into the 
above equation gives
\begin{equation}
E_{nl} \approx 2 \gamma ( 2 n + l + 3/2) + \frac{\gamma^{1/2}}{2^{1/2}} 
\sum_{k,k'=0}^n a_{nl,2k} a_{nl,2k'} \Gamma(l+k+k'+1) {\cal N}_{nl}^2 
\label{eq16}
\end{equation}
where $a_{nl,2k}$ are the coefficients of the generalized Laguerre 
polynomials (of Eq.(\ref{eq4})) that satisfy the recursion
\begin{equation}
a_{nl,2k} = a_{nl,2(k-1)} \frac{k-n-1}{k(k+l+1/2)}
\label{eq17}
\end{equation}
with $a_{nl,0}=1$. Furthermore, the normalization constant is given by
\begin{equation}
{\cal N}_{nl}^{-2} = \sum_{k,k'=0}^n a_{nl,2k} a_{nl,2k'} 
\Gamma(l+k+k'+3/2) \; .
\label{eq18}
\end{equation}

Based on the perturbative results of Eqs. (\ref{eq13b}), (\ref{eq14}) 
and (\ref{eq16}) we propose the following interpolation formula
\begin{equation}
E_{nl}^{(int)}(\gamma) = \frac{\gamma^{-1} E_{nl}^{(0)}+
\gamma E_{nl}^{(\infty)}}{\gamma^{-1}+\gamma}  \; ,
\label{eq19}
\end{equation}
where the superscripts $(0)$ and $(\infty)$ correspond to the zero and 
infinite confinement limits. For $\gamma \rightarrow 0$ equation (\ref{eq19})
 will return
approximately the weak confinement limit and for $\gamma \rightarrow 
\infty$ the strong confinement limit. For the zeroth order approximations
equation (\ref{eq19}) leads to
\begin{equation}
E_{nl}^{(int)}(\gamma) = \frac{ \gamma^{-1} U_0^l + 2\gamma^2 (2n+l+3/2)}{
\gamma^{-1}+\gamma} \; ,
\label{eq20}
\end{equation}
and for the first order pertubative results, it will be
\begin{equation}
E_{nl}^{(int)}(\gamma) = \frac{ \gamma^{-1}[ U_0^l + 2 \gamma_l ( n + 1/2 )] 
+ \gamma [ 2 \gamma (2n+l+3/2) + 
 2^{-1/2} \gamma^{1/2} \Delta_{nl} ]}{ \gamma^{-1}+\gamma} \; ,
\label{eq21}
\end{equation}
where $\Delta_{nl}$ denotes the summation in the right hand side of equation 
(\ref{eq16}). The interpolation scheme of equation (\ref{eq20}) performs
consistently bad, except for the very weak, and very strong fields. The
error goes up to $41\%$ for the ground-state, underestimating the correct 
values. The introduction of first order corrections, corresponding to 
the interpolation scheme of equation (\ref{eq21}) brings a dramatic 
improvement on the values: for the ground-state the relative error is never
worse than $3.3\%$ for the energy of the relative motion, which means that 
the relative error of the total energy is around $1.7\%$.

In the case of the mean-field approximations like Hartree-Fock and
Kohn-Sham, we have considered the paired-electron ground-state case, 
and since the confining potential is radially symmetric, the orbitals 
and the density are also radially symmetric. We implemented the 
shifted-1/N technique for the HF and KS equations. 
Here, a word about accuracy is needed: although the solution of the
Schr\"odinger-like equations using Schwartz's method is very accurate,
the estimate of the Hartree potential is not as accurate anymore, 
nevertheless, five to six figures are always guaranteed. 
The results are 
shown in table 1 in the first and third columns, indicated as $HF-1/N$
and $KS-1/N$, respectively. Since the
resulting wavefunction from the application of the 1/N has a 
complicated form which makes difficult a direct evaluation of 
the density or of the Hartree potential, we assumed that as initial
guess the non-interacting density, which is also correct asymptotically.
For comparison purposes, we have used again the Schwartz technique 
self-consistently to solve both the HF and KS equations. Numerical 
results are shown in the second and fourth columns of table 1,
indicated as $HF-num$ and $KS-num$, respectively.
Also for comparison purposes, we included the results of
solving the full Schr\"odinger equation using the shifted-1/N 
method and the numeric solution, which are the last two columns
indicated as $Exact-1/N$ and $Exact-num$, respectively. 

Table 1 shows very good agreement between the results from the accurate
numerical result and the ones calculated with the 1/N method. We should 
not be too enthusiastic about the accuracy, since the remarkable agreement 
is probably a result of the compensation of errors from the calculation of
the Hartree potential and energy (due to its simplicity) and the intrinsic
error of the 1/N method, specially for the weak field case. It can also 
be seen that the relative accuracy improves from typically few times 
$10^{-2}$ to $10^{-4}-10^{-5}$ with the increase of the strength
of the field. Also the systematic difference between the HF and KS
methods is reduced with stronger confinement. Both behaviors can be
understood by taking into account that the confinement potential 
dominates over the decaying electron-electron interaction potential,
and with the increase of the strength of confinement the problem becomes
just a harmonic potential problem, for which the shifted-1/N is designed 
to give the exact energy, although the quality of the wavefunction is not
too good.

In order to better understand the behavior of the pair of electrons in
strong and weak confinement within these mean-field theories, we can 
recourse to pertubation analysis. In the strong 
confinement limit ($\gamma \rightarrow \infty$) the system will behave 
basically as a pair of non-interacting electrons, and the electron-electron
interaction (Hartree and exchange potentials) can be considered as a 
perturbation. The HF and KS orbitals become
\begin{equation}
\phi(\vec r) = \left(\frac{\gamma}{\pi} \right)^{3/4} \exp (- \frac 12 
\gamma r^2) \; ,
\label{eq22}
\end{equation}
so, the Hartree energy will be $2 (2\gamma/\pi)^{1/2}$ and the LDA
exchange $3 c_x 2^{1/3} (3\gamma/\pi)^{1/2}/4$. Then the total HF
energy will be
\begin{equation}
E_{HF}^{(\infty)} = 3 \gamma + \left( \frac{2 \gamma}{\pi} \right)^{1/2} \; ,
\label{eq23}
\end{equation}
and the total KS energy
\begin{equation}
E_{KS}^{(\infty)} = 3 \gamma + \left( 2^{3/2}+\frac 34 2^{1/3} 3^{1/2} c_x 
\right) \left( \frac{\gamma}{\pi} \right)^{1/2} \; .
\label{eq24}
\end{equation}
In the weak confinement limit ($\gamma \rightarrow 0$), 
due to scaling arguments, we can neglect
the kinetic energy, and we can assume constant density, at least within
a certain radius $R$ (this is asymptotically true for the HF approximation
and arguably for the KS one). Now taking Eqs. (\ref{hf1}) and 
(\ref{ks1}), and using Poisson's equation,  we get that $\rho_{HF} 
\approx \gamma^2/ (2 \pi)$ and
$\rho_{KS} \approx \gamma^2/ (4 \pi)$. Then from normalization
condition $R = [ 3/( 2\pi \rho) ]^{1/3}$. The Hartree potential will
take the form
\begin{equation}
v_H(r) = \frac 3R - \frac{r^2}{R^3} \; , \qquad if \; r \leq R \; ,
\label{eq25}
\end{equation}
and $v_H(r)=2/r$ if $r \geq R$. The Hartree energy will be equal to
$12/(5 R)$. Substituting the above result into equation (\ref{hf2}) we 
get that the Hartree-Fock energy will be
\begin{equation}
E_{HF}^{(0)} \approx (3 \gamma)^{2/3}  \; .
\label{eq26}
\end{equation}
Substituting the result for the Hartree energy into equation (\ref{ks2}) 
we find that in this limit the Kohn-Sham energy will be equal to
\begin{equation}
E_{KS}^{(0)} \approx \gamma^{2/3} ( 6^{3/2} + \frac{2 c_x}{5 (4\pi)^{1/3}})  
\; .
\label{eq27}
\end{equation}
Again, we can use interpolation schemes like the one of equation 
(\ref{eq19}), for the HF ground-state energy we will have
\begin{equation}
E_{HF}^{(int)}(\gamma)  \approx \frac{(3 \gamma)^{2/3}+ \gamma^2 
[ 3 \gamma + ( 2 \gamma/ \pi)^{1/2} ] }{1+\gamma^2}  \; .
\label{eq28}
\end{equation}
In the Kohn-Sham case we can write down
\begin{equation}
E_{KS}^{(int)}(\gamma) \approx \frac{ \gamma^{2/3} [ 6^{3/2} + 
2 c_x/5 /(4\pi)^{1/3} ] + \gamma^2 [ 3 \gamma + ( 2^{3/2} + 
3 \; 2^{1/3} 3^{1/2} c_x/4 ) ( \gamma/\pi )^{1/2} ] }{1+\gamma^2}  \; .
\label{eq29}
\end{equation}

Comparing the asymptotics of the exact equation for the ground-state in 
the strong confinement limit with the asymptotics of the HF equation, we 
observe that they coincide. In the Kohn-Sham case the zeroth order is the
same, but the coefficient of the first order perturbation is slightly
higher ($1.62$ compared to the exact $\sqrt{2} \approx 1.41$). In the low
density limit (weak confinement), although the three asymptotes are
proportional to $\gamma^{2/3}$ in the zeroth order, the coefficients differ 
substantially, the exact is $1.19$, the HF $2.08$, and the KS $3.17$.
This indicates a consistent overestimation of the energy by the 
independent-electron approximations, which is the expected behavior at
least for the HF approximation. Our findings for this system for the
weak and strong asymptotae, and the numerical results, suggest that
KS-LDA energies are always higher than the HF ones, which is consistent 
with numerical experience on atoms and molecules \cite{PY}.
From the results shown in table 1 we can estimate the correlation energies
for the ground-state 
for different confinement strengths: for $\gamma^{\; \prime}=0.1$, the
correlation energy is about $5.6\%$ of the total energy, meanwhile for
$\gamma^{\; \prime}=0.9$ (strong confinement) it is about  $0.17\%$
(for the lightest many-electron atom, Helium, it is about $1.4\%$, and 
for Argon with
$Z=18$, it is only $0.14\%$). This adds evidence to the suggestion that for
harmonic fields correlation effects are more important, also in three
dimensions, especially for weak confinement.

\section{Concluding remarks}

In the present article we have calculated  the energy spectrum of a 
two-electron 
spherical quantum dot for the few lowest states, using the shifted-1/N 
method and the very accurate numerical Schwartz's method. From the
comparison of the numerical results we could assess the quality of the 
shifted-1/N method, which consistently underestimates the correct values,
although the error is rather small, and it decreases with the increase
of the relative angular momentum. We have also applied perturbation theory
up to first order in both limits of confinement of the electron, and we
have proposed some interpolation formulae for the energy. Inclusion of
first order perturbation allowed to construct an interpolation expression
that performs reasonably well.  We also solved the mean-field Hartree-Fock
and local-density Kohn-Sham problems for the ground-state. 
Using the shifted-1/N method we got from reasonable to high accuracy 
already in the first iteration, compared to the self-consistent numeric 
solution using the Schwartz's method. We made an analysis of the strong 
and weak confinement limits, and proposed interpolation formulae for both
the Hartree-Fock and Kohn-Sham ground-state energies. It was shown that
the correlation energy is relatively big for this systems, especially 
for weak confinement.

\section{Acknowledgments}

One of the authors (V.M.V.) would like to acknowledge support by CONICIT,
Venezuela, under project 96000061.



\begin{figure}
  \vspace*{18cm}
  \hspace*{-3cm}
  \includegraphics{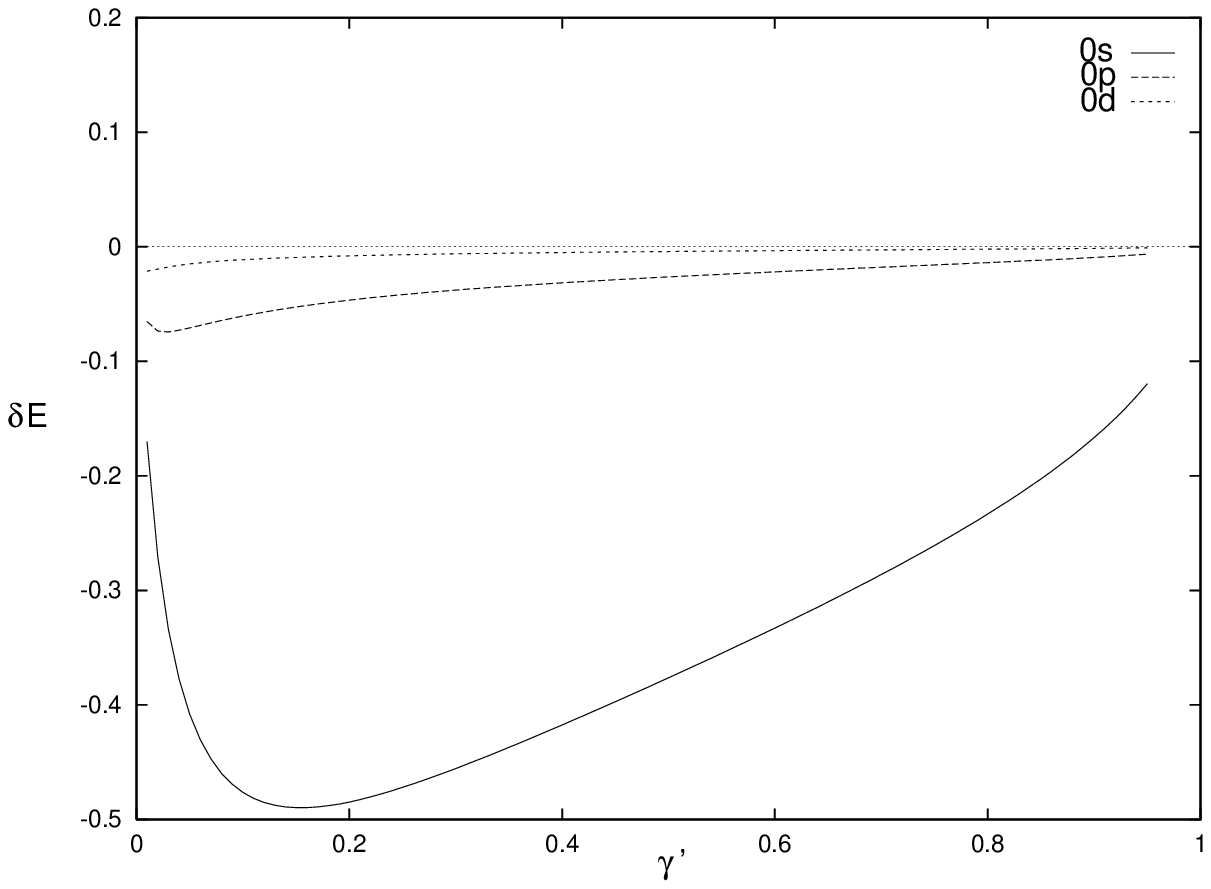}
\caption{Relative error ($\delta E = (E_{1/N}/E_{exact}-1) \times 100$) of 
the energy of the relative motion for the lowest energy states for $l=0, 1$, 
and $2$ calculated with the 1/N method, as a function of the reduced 
frequency $\gamma^{\; \prime}=\gamma /(\gamma +1)$.}
\end{figure}

\newpage
\begin{figure}
  \vspace*{18cm}
  \hspace*{-3cm}
  \includegraphics{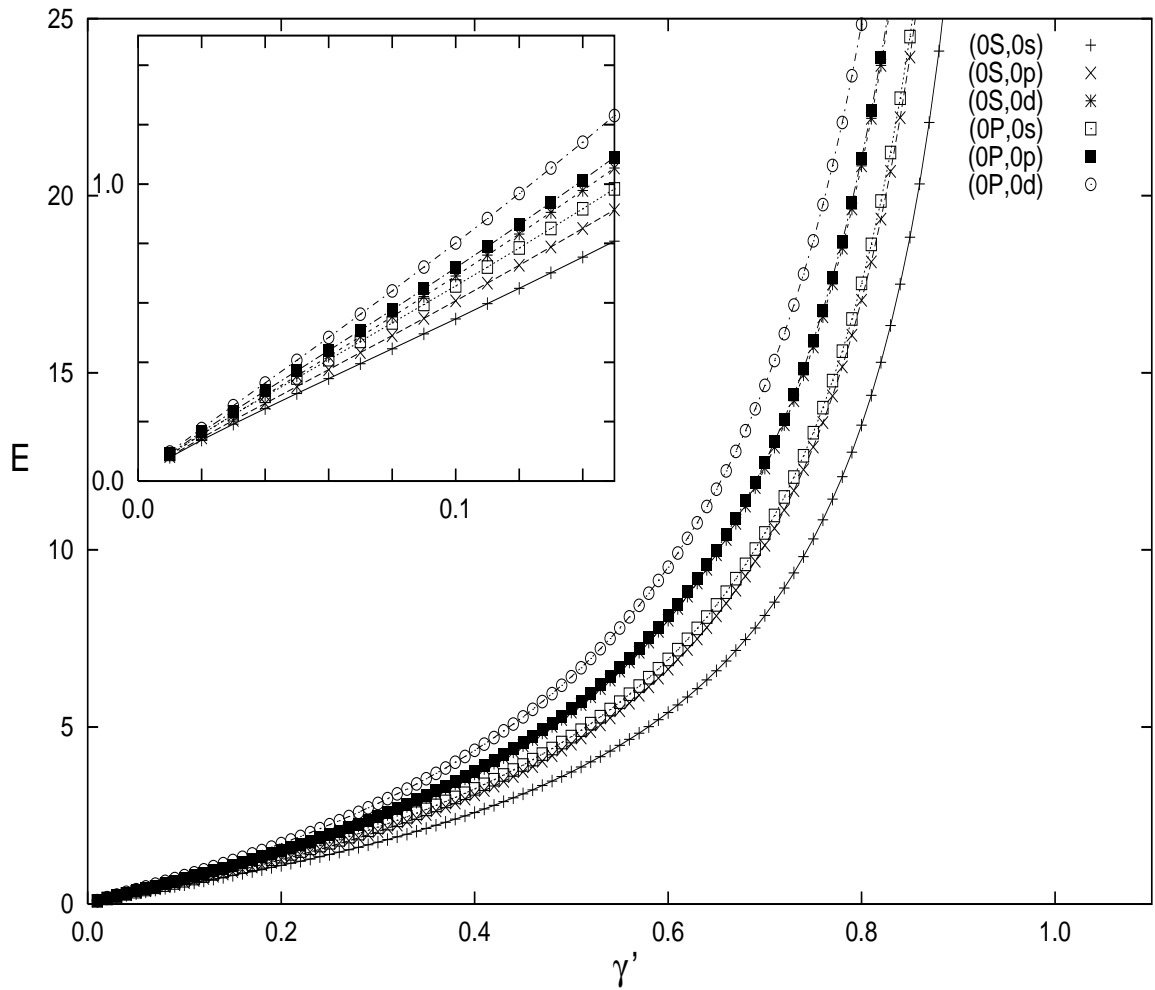}
\caption{Total energy as a function of $\gamma^{\; \prime}=\gamma 
/(\gamma +1)$ for the few lowest energy states. $0S$, and $0P$ depict the 
lowest energy states for the centre-of-mass motion for $L=0$, and $1$, 
respectively, and $0s, 0p$, and $0d$ the lowest energy states for the 
relative motion for $l=0, 1$, and $2$, respectively. The inset shows a 
magnification for small $\gamma^{\; \prime}$. }
\end{figure}


\begin{table}[h]
\begin{center}
\begin{tabular}{|c|c|c|c|c|c|c|}
$ \;\; \gamma^{\; \prime} \;\; $ & $HF-1/N $ & $HF-num  $ & 
$KS-1/N  $ & $KS-num  $ & $Exact-1/N $ & $Exact-num $\\
\hline
  0.1  & 0.5666 & 0.5768 & 0.5960 & 0.6082 & 0.5443 & 0.54606 \\
\hline
  0.2  & 1.1163 & 1.1241 & 1.1644 & 1.1742 & 1.0858 & 1.08926 \\
\hline
  0.3  & 1.7758 & 1.7826 & 1.8408 & 1.8503 & 1.7398 & 1.74478 \\
\hline
  0.4  & 2.6200 & 2.6255 & 2.7029 & 2.7118 & 2.5791 & 2.58569 \\
\hline
  0.5  & 3.7673 & 3.7717 & 3.8711 & 3.8791 & 3.7217 & 3.73012 \\
\hline
  0.6  & 5.4477 & 5.4508 & 5.5775 & 5.5842 & 5.3972 & 5.40775 \\
\hline
  0.7  & 8.1906 & 8.1922 & 8.3558 & 8.3608 & 8.1345 & 8.14778 \\
\hline
  0.8  & 13.5693 & 13.5693 & 13.7902 & 13.7928 & 13.5057 & 13.5232 \\
\hline
  0.9  & 29.3703 & 29.3679 & 29.7094 & 29.7082 & 29.2930 & 29.3194  
\end{tabular}
\end{center}
\caption{Three-dimensional two-electron quantum dot total energy, using the
Hartree-Fock ($HF$), Kohn-Sham ($KS$), and exact Schr\"odinger ($Exact$) 
equations for selected values of the reduced confinement constant 
($\gamma^{\prime}$), calculated using the shifted-1/N ($1/N$), and Schwartz's 
numeric ($num$) methods.}
\end{table}

\end{document}